\documentclass{article}
\usepackage{spconfa4,amsmath,graphicx,url,times,subcaption,bm,siunitx,float}

\usepackage{pgfplots}
\usepackage{color}
\copyrightnotice{978-1-5386-8151-0/18/\$31.00~\copyright2018 IEEE}

\mathchardef\mhyphen="2D %

\newcommand\norm[1]{\left\lVert#1\right\rVert}

\title{exploring tradeoffs in models for low-latency speech enhancement}
\name{Kevin Wilson, Michael Chinen, Jeremy Thorpe, Brian Patton, }
\secondlinename{John Hershey, Rif A. Saurous, Jan Skoglund, Richard F. Lyon }
\address{Google Research, Google Chrome Audio}
\begin{document}
\maketitle

\begin{abstract}
We explore a variety of neural networks configurations for one- and
two-channel spectrogram-mask-based speech enhancement. Our best model improves on
previous state-of-the-art performance on the CHiME2 speech enhancement task by 0.4 decibels in signal-to-distortion ratio (SDR).
We examine trade-offs such as non-causal look-ahead, computation, and parameter count versus enhancement performance and find that zero-look-ahead models can achieve, on average, within 0.03 dB SDR of our best bidirectional model.  Further, we find that 200 milliseconds of look-ahead is sufficient to achieve equivalent performance to our best bidirectional model.
\end{abstract}

\section{Introduction}
\label{sec:intro}
Recent work \cite{williamson2016complex,erdogan2015phase,chen2017thesis,chen2016large,chen2016long,weninger2015speech} has successfully used deep learning to train systems to estimate multiplicative spectrogram masks to remove noise from mixtures of speech and noise.  In this work, we explore the design space of such systems in the context of the CHiME2 WSJ0 dataset \cite{vincent2013second} by varying the input features, the loss function, the representation of the mask, and the size and shape of the trainable neural network.

To explore this space, we first search the space of fully bidirectional (non-causal) models using Vizier \cite{golovin2017vizier}, Google's hyperparameter tuning system.  Then, starting with 
the best-performing configuration found by Vizier, we examine the effect of look-ahead on enhancement performance.

\begin{figure*}[!h]
  \centering
  \centerline{\includegraphics[width=\textwidth,clip]{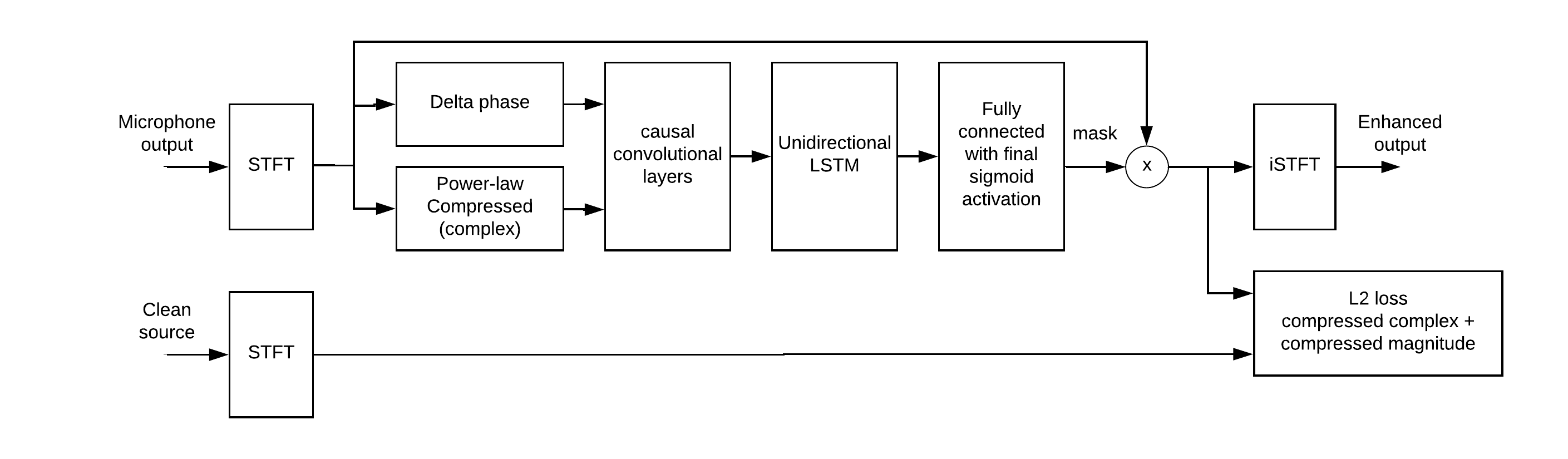}}
  \caption{System architecture.}
  \label{fig:architecture_block_diagram}
\end{figure*}

\section{Background}

Our speech enhancement networks are similar to those used in other recent work on spectrogram-mask-based speech enhancement and are shown schematically in Figure \ref{fig:architecture_block_diagram}.  They take as input a noisy, possibly multi-microphone, time-domain audio signal.  We then compute the short-time Fourier transform (STFT), and from the STFT we derive input features, e.g. the compressed magnitude spectrogram, for the network.  Our neural networks consist of a stack of convolutional layers followed, optionally, by LSTM and fully connected layers, each with ReLU activations.  The final layer has a sigmoidal activation and outputs a soft mask, which is pointwise multiplied by the noisy input STFT.  During training, this masked STFT is compared to the clean reference STFT according to a loss function (described below).  To apply the model, the inverse STFT of the masked STFT is computed to generate an enhanced time-domain output.

In this work, we use the CHiME2 WSJ0 dataset \cite{vincent2013second}, which consists of cleanly-recorded Wall Street Journal sentences convolved with a standard set of two-microphone room impulse responses and added to recorded two-microphone noise.  The signal-to-noise ratios (SNRs) of the resulting utterances range from -6 to 9 dB.  In this dataset, the target speaker is always at broadside (no relative delay between the two microphones), and the training target signals have been passed through the room impulse responses, so the goal is to remove additive noise only, not to dereverberate the signal.

The previous state-of-the-art enhancement performance on CHiME2 WSJ0 was achieved by Weninger et al. \cite{weninger2015speech}.  Their system consisted of 2 bidirectional LSTM layers, each with 384 units, followed by a fully connected layer with sigmoidal activation.  Weninger et al.\ use a loss function, ``phase-sensitive spectrum approximation,'' which modifies only the magnitudes (not the phases) of the noisy input, but which penalizes the output based on distance in the complex spectrum space.  
Their best system also uses speech recognizer state estimates as additional inputs.  It achieves 15.07 dB output SDR averaged over the evaluation dataset.

We do not attempt to replicate all details of their configurations, but we take their system as inspiration and, by doing a large hyperparameter search, find a system that improves on their system's performance.

\section{Model and training}

Starting with 16 kHz sample rate audio, we compute the STFT using a 25 ms Hann window and a 10 ms hop size, resulting in a 257-dimensional STFT frame every 10 ms.  We use this same STFT as the basis for our loss function, the input to our neural network, and as the representation to which the time-frequency mask is applied.  The loss function for our models is
\begin{equation}
\begin{split}
  \label{eqn:loss}
  L  = & \norm{ |S_{clean_0}^{0.3}(t,f)| - |S_{enhanced}^{0.3}(t,f)|}_2^2 + \\
  & \lambda \norm{S_{clean_0}^{0.3}(t,f) - S_{enhanced}^{0.3}(t,f)}_2^2
\end{split}
\end{equation}
where $S$ is an STFT representation.  The first component penalizes mismatches between the enhanced {\em magnitude} spectra and channel 0 of the clean training target, and the second component penalizes mismatches between the enhanced {\em complex} spectra and channel 0 of the training target, weighted by hyperparameter $\lambda$.  All spectra are power-law compressed (with power $0.3$), which partially equalizes the importance of quieter sounds relative to loud ones.  If we had set the power in our power-law compression to 1.0, the second term of our loss function would be equivalent to the phase-sensitive spectrum approximation (PSA) loss from Weninger et al. \cite{weninger2015speech}.

We train on batches of fixed-length (3 second) audio clips, obtained by chopping the variable-length CHiME2 training examples into consecutive 3-second chunks and discarding any final fractional chunks. 

\section{Vizier search}

There are many possible variations on the high-level architecture described above.  We implement our models in TensorFlow and use Vizier \cite{golovin2017vizier} to search this space using its Batched Gaussian Process Bandits approach.  In early experiments, we observed that models usually achieved their maximum validation set performance within 24 hours on one GPU, while consuming one to six training epochs depending on the complexity of the model.  For our Vizier exploration, we trained each model for up to 24 hours, using maximum validation set performance as Vizier's objective function.  At each stage, Vizier suggested parameters for 40 models that were trained concurrently, until a total of 765 models were trained.

\begin{table}[ht]
  \centering
  \begin{subtable}{\columnwidth}
  \centering
  \begin{tabular}{|c|c|c|c|c|}
    \hline
    Filters & T width & F width & T dilation & F dilation \\
    \hline
          32 & 1 & 7 & 1 & 1 \\
          32 & 7 & 1 & 1 & 1 \\
          32 & 5 & 5 & 1 & 1 \\
          32 & 5 & 5 & 2 & 1 \\
          32 & 5 & 5 & 4 & 1 \\
          32 & 5 & 5 & 8 & 1 \\
          32 & 5 & 5 & 16 & 1 \\
          8 & 1 & 1 & 1 & 1 \\    
    \hline
  \end{tabular}
  \caption{Small configuration}
  \end{subtable}
  \begin{subtable}{\columnwidth}
  \centering
  \begin{tabular}{|c|c|c|c|c|}
    \hline
    Filters & T width & F width & T dilation & F dilation \\
    \hline
          32 & 1 & 7 &  1 & 1 \\
          32 & 7 & 1 &  1 & 1 \\
          32 & 5 & 5 &  1 & 1 \\
          32 & 5 & 5 &  2 & 1 \\
          32 & 5 & 5 &  4 & 1 \\
          32 & 5 & 5 &  8 & 1 \\
          32 & 5 & 5 & 16 & 1 \\
          32 & 5 & 5 & 32 & 1 \\
          32 & 5 & 5 &  1 & 1 \\
          32 & 5 & 5 &  2 & 2 \\
          32 & 5 & 5 &  4 & 4 \\
          32 & 5 & 5 &  8 & 8 \\
          32 & 5 & 5 & 16 & 16 \\
          32 & 5 & 5 & 32 & 32 \\
          8 & 1 & 1 & 1 & 1 \\
    \hline
  \end{tabular}
  \caption{Large configuration}
  \end{subtable}
  \caption{Convolutional layer configurations explored in our Vizier study. ``Filters'' is the number of filters (feature maps) in the layer.  ``T width'' and ``F width'' are the size of the filter in time (frames) and frequency (bins), respectively.  ``T dilation'' and ``F dilation'' are dilation factors in time and frequency, respectively.}
  \label{table:conv}
\end{table}

\begin{table}[ht]
  \centering
  \begin{tabular}{|c|l|}
    \hline
    Hyperparameter & Possible values \\
    \hline
    Convolutional config & small, large (See Table \ref{table:conv}.) \\
    BLSTM depth & $ 0 \mhyphen 5 $ \\
    BLSTM width & $ 8 \mhyphen 1024 $ \\
    Fully connected depth & $ 0 \mhyphen 5 $ \\
    Fully connected width & $ 8 \mhyphen 1024 $ \\
    Delta-phase input & yes, no \\
    Complex loss $\lambda$ & $  0.0 \mhyphen 1.0 $ \\
    Learning rate & $ \num{3e-6} \mhyphen \num{1e-3} $ \\
    Input channels & $ 1 \mhyphen 2$ \\ 
    \hline
  \end{tabular}
  \caption{Hyperparameters explored in our Vizier study. }
  \label{table:hyperparameters}
\end{table}

All of our networks use power-law compressed magnitude spectrograms as input features.  For ``delta-phase'' models, we concatenate delta-phase spectrogram inputs \cite{mccowan2011delta}, which are frame-to-frame phase ratios at each frequency.  Two-channel inputs yield two-channel spectrogram images, so using delta-phase doubles the channel dimension.%

For the neural network, we stack one of the two configurations of convolutional layers described in Table \ref{table:conv}, followed by LSTM layers with residual connections (where each layer is put in parallel with a bypass connection), followed by fully connected layers.  Preliminary experiments showed improvement from having some convolutional layers, but it was not feasible to do a less-constrained search over convolutional architectures while also exploring other aspects of the model.  Thus, we limit this study to these two possible convolutional layer configurations.

All models in this study use the output of the DNN as a real-valued mask by which the input signal(s) are multiplied to produce denoised spectrogram.  In the case of two-channel input, that results in adding together the magnitude-masked channels, with no relative delay.  Since the target speaker in CHiME2 is always at zero relative delay, this is reasonable, and preliminary experiments in which we allowed the network to adjust the relative phase yielded no benefit.  For tasks in which the direction of the target varies, we would expect the optimal mask to have non-zero phase in general.

We evaluate model accuracy and select the best performing model using source-to-distortion ratio (SDR) from BSS Eval \cite{vincent2006performance}. 
While this metric is imperfect (it allows the enhancement output to be badly equalized), we use it because it is a standard benchmark.  By this metric, our best model found by the search achieved an average development set SDR of 14.60 dB and eval set SDR of 15.37 dB (Table \ref{table:best_sdr_bidi}).

Our best model includes delta-phase input and uses relatively wide BLSTM and fully connected layers, but it uses the smaller of the two convolutional configurations.  It was trained using a loss function that includes both magnitude and complex loss, with Vizier finding that a small amount of complex-loss, $\lambda = 0.113$, was optimal.

The hyperparameter values and ranges over which Vizier searched are in Table \ref{table:hyperparameters} and the hyperparameter values for the best model found by Vizier are in Table \ref{table:best_model}.

\begin{table}[h]
  \centering
  \begin{tabular}{|c|l|}
    \hline
    Hyperparameter & Value \\
    \hline
    Convolutional config & small \\
    BLSTM depth & $ 3 $ \\
    BLSTM width & $ 1023 $ \\
    Fully connected depth & $ 2 $ \\
    Fully connected width & $ 873 $ \\
    Delta-phase input & yes \\
    Complex loss $\lambda$ & $  0.113 $ \\
    Learning rate & $ \num{2.1e-4} $ \\
    Input channels & $ 2 $ \\ 
    \hline
  \end{tabular}
  \caption{Best-performing hyperparameters found in our Vizier study. }
  \label{table:best_model}
\end{table}

\begin{table}[!tbh]
  \centering
  \begin{tabular}{|c|c|c|c|c|c|c|}
    \hline
    \multicolumn{7}{|c|}{Eval SDR} \\
    \hline
    -6dB & -3dB & 0dB & 3dB & 6dB & 9dB & Avg \\
    \hline
          12.17 & 13.44 & 14.70 & 15.83 & 17.30 & 18.78 & 15.37 \\  
    \hline
  \end{tabular}
  \caption{Performance of best-performing model found by Vizier search as a function of input SNR.}
  \label{table:best_sdr_bidi}
\end{table}

\begin{table}[!tbh]
  \centering
  \begin{tabular}{|c|c|c|c|c|c|c|}
    \hline
    \multicolumn{7}{|c|}{Eval SDR} \\
    \hline
    -6dB & -3dB & 0dB & 3dB & 6dB & 9dB & Avg \\
    \hline
          12.31 & 13.52 & 14.76 & 15.94 & 17.41 & 18.90 & 15.48 \\  
    \hline
  \end{tabular}
  \caption{Performance of best-performing causal model (180 ms look-ahead) as a function of input SNR.}
  \label{table:best_sdr_lookahead}
\end{table}

We compared SDR, number of weights, and operations per second of audio processed of the 765 models that were explored in the Vizier search.  In general, we found that more LSTM and fully-connected weights improved SDR, with diminishing returns, for the small convolutional network.  The large convolutional network required approximately twice the operations, and may not have converged by the 24 hours deadline with large LSTM/fully-connected layers.

\begin{figure}[!thbp]
  \centering
  \begin{subfigure}{\columnwidth}
    \centering
    \includegraphics[width=\textwidth,height=1.7in]{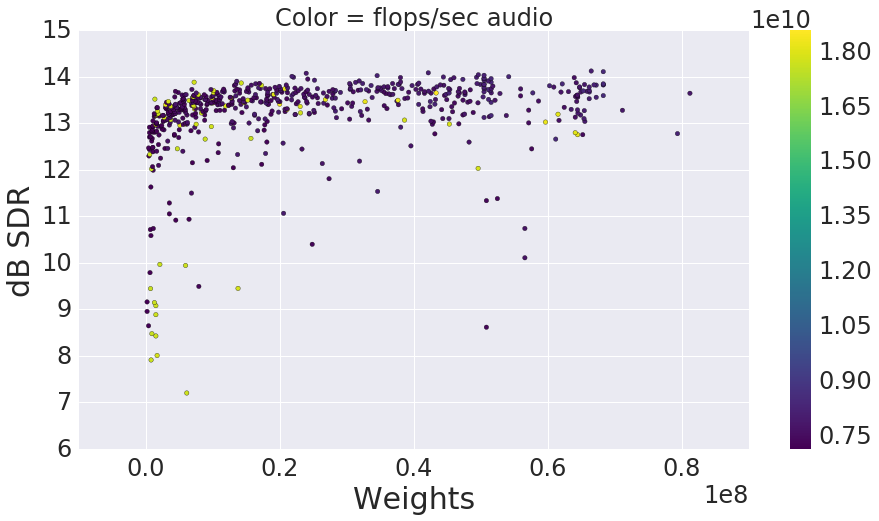}
    \caption{SDR vs. number of weights for all models in our Vizier study.}
    \label{fig:weights_vs_sdr}
  \end{subfigure}
  \begin{subfigure}{\columnwidth}
    \centering
    \includegraphics[width=\textwidth,height=1.7in]{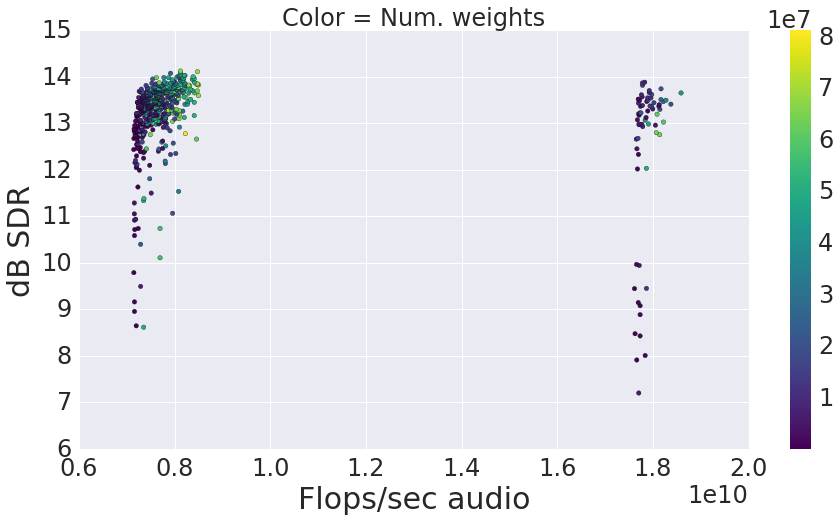}
    \caption{SDR vs. operations  per second of audio for all models in our Vizier study. The two clusters reflect the two configurations of convolutional layers, small and large, defined in Table \ref{table:conv}.}
    \label{fig:flops_vs_sdr}
  \end{subfigure}
\caption{Scatter plots of the models trained in our Vizier study.}
\label{fig:vizier_scatter}
\end{figure}

In Figure \ref{fig:weights_vs_sdr}, we show the maximum SDR achieved on the development set vs. number of trainable weights.  While the top performing model has around 65 million parameters, the Vizier study found models with only around 1M parameters that perform within about 1dB.  Vizier seems to have explored the space of models with many/few parameters fairly well.

In the set of models that we explored, computation was dominated by the convolutional layers, as can be seen in figure \ref{fig:flops_vs_sdr}.  The number of operations per second required by the two convolutional configurations is approximately equal to the minimum value in each of the two clusters.  All of the models we explored have a fairly high inference-time computational cost.  A finer-grained class of convolutional layers, and possibly a different optimization metric, would be needed to find models appropriate for tight computational budgets.

The SDR numbers in Figure \ref{fig:vizier_scatter} are lower than the results in other tables and figures because the Vizier objective scores used for Figure \ref{fig:vizier_scatter} use only a subset of the CHiME2 development set, which is more difficult than the evaluation set.

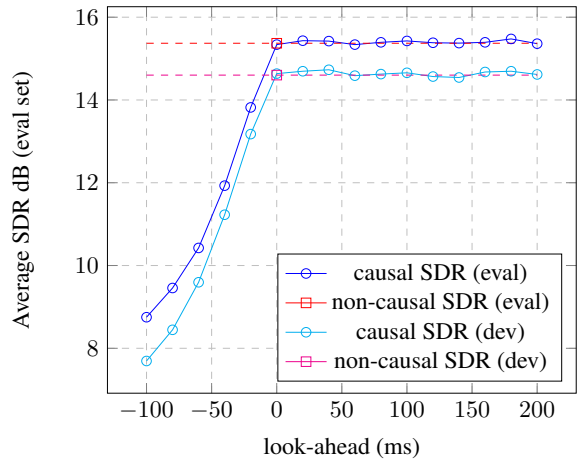
\begin{figure}[h]
\centering
    \begin{tikzpicture}[scale=0.9]
      
      \begin{axis}[
        xlabel={look-ahead (ms)},
        ylabel={Average SDR dB (eval set)},
        grid style=dashed,
        xmajorgrids=true,
        ymajorgrids=true,
        legend pos=south east
      ]
      \addplot[
        color=blue,
        mark=o,
      ]
      coordinates {
        (-100, 8.750622192)
        (-80, 9.45675314)
        (-60, 10.42473717)
        (-40, 11.9295185)
        (-20, 13.82000407)
        (0, 15.33760053)
        (20, 15.43260096)
        (40, 15.42226718)
        (60, 15.33770566)
        (80, 15.39292041)
        (100, 15.42662311)
        (120, 15.38269229)
        (140, 15.37382213)
        (160, 15.39288302)
        (180, 15.47509701)
        (200, 15.36033731)
      };
      \addlegendentry{causal SDR (eval)}
      \addplot[
        color=red,
        mark=square,
      ]
      coordinates {
        (0, 15.37063314)
      };
      \addlegendentry{non-causal SDR (eval)}
      
      \addplot[
        color=cyan,
        mark=o,
      ]
      coordinates {
        (-100, 7.693090085)
        (-80, 8.446376102)
        (-60, 9.595072993)
        (-40, 11.2271429)
        (-20, 13.17532223)
        (0, 14.6365332)
        (20, 14.6941872)
        (40, 14.72992517)
        (60, 14.58711172)
        (80, 14.62505575)
        (100, 14.65653263)
        (120, 14.56607141)
        (140, 14.54316881)
        (160, 14.67587566)
        (180, 14.69447552)
        (200, 14.61635616)
      };
      \addlegendentry{causal SDR (dev)}
    
      \addplot[
        color=magenta,
        mark=square,
      ]
      coordinates {
        (0, 14.60063708)
      };
      \addlegendentry{non-causal SDR (dev)}   
      
      \addplot[
        color=red,
        style=dashed,
        domain=-100:200
      ]
      {15.37063314};

      \addplot[
        color=magenta,
        style=dashed,
        domain=-100:200
      ]
      {14.60063708};

      \end{axis}
      
    \end{tikzpicture}
    \caption{Plot of SDR vs. look-ahead }
    \label{fig:lookahead}
\end{figure}

\section{Effects of look-ahead}

We create a causal variant of the best configuration found by Vizier by changing all LSTM layers from bidirectional to unidirectional (which cuts the number of parameters per layer in half) and by altering the receptive field of each convolutional layer to be causal rather than centered on the current time (which does not change the number of parameters per layer).  With that causal configuration, we then shift the input features with respect to the output, yielding a varying amount of future context while keeping the number of parameters fixed, and examine the effect of look-ahead on denoising performance.  To control for the run-to-run variance we ran 16 trials of each look-ahead setting, picked the model and training step that best performed on a subset of the development set, and scored it against the evaluation dataset (Table \ref{table:best_sdr_lookahead}).  The scores for the non-causal model were also found this way.

Our findings, which are consistent with Wichern and Lukin \cite{wichern2017low} and Erdogan et al.\ \cite{erdogan2015phase}, but which more systematically explore look-aheads, are that unidirectionality with zero or positive look-ahead shows little difference in SDR compared to the bidirectional model for the CHiME2 dataset.  For non-negative look-ahead, the differences in SDR had a range of only 0.14dB.  Such differences are of marginal statistical significance based on our finding a run-to-run standard deviation of 0.08dB SDR for multiple trainings of identical configurations.  For negative look-ahead (predicting future spectrogram mask values), the effect was much larger, with a reduction of 6.6dB SDR with -100 ms look-ahead compared to the zero-look-ahead causal model.  When the network must blindly predict future masks, it will not be able to respond immediately to transient in the noise, or to the nonstationarity of the speech.  In contrast, positive look-ahead may have less of an effect because of the predictability of clean speech. 

Figure \ref{fig:lookahead} shows an unexpectedly small gain in performance as look-ahead increases (among non-negative look-ahead values).  This contrasts with results in \cite{wichern2017low} as well as our unpublished experiments using single-channel inputs, for which the beneficial effect of look-ahead was larger. One hypothesis is that two-channel models 
have a larger input dimensionality, and hence a greater capacity to overfit. 
In addition, CHiME2 training/validation/test datasets are constructed using binaural impulse responses, recorded under different conditions (e.g. doors and curtains open/closed, different ambient noise).  Thus there may be significant mismatch in the spatial characteristics between training and validation data, leading to a tendency for overfitting with two-channel models.  In preliminary experiments with dropout regularization we observed improved performance for large look-ahead models, presumably via a reduction in overfitting.  

\section{Conclusion}

We described a spectrogram-mask-based speech enhancement system, which takes as input the noisy magnitude spectrogram and delta-phase spectrogram, and consists of a stack of convolutional, LSTM, and fully connected neural network layers, that achieves a new state-of-the-art performance on the CHiME2 speech enhancement dataset.  We found the cost of unidirectionality to be negligible with zero or positive look-ahead, and large with negative look-ahead.

\bibliographystyle{IEEEbib}
\bibliography{refs}

\end{document}